\documentclass[a4paper,12pt]{article}
\usepackage[dvips]{graphicx}
\DeclareGraphicsExtensions{.pdf,.jpg}

\usepackage{amsfonts,amsmath,amsthm,amssymb}
\usepackage{mathrsfs}
\usepackage[onehalfspacing]{setspace}
\author{L. Ort\'{i}z\footnote{leonardo.ortiz@uniofyorkspace.net}}

\title{The Energy-Momentum Tensor in the 1+1 dimensional non-rotating BTZ black hole}

\begin{document}

\maketitle

\begin{center}



Department of Mathematics\\
The University of York\\
York YO10 5DD, U. K.\\\vspace{0.4cm}

\normalsize{\textbf{Abstract}}\\\end{center} \small{We study the
energy-momentum tensor for the real scalar field on the 1+1
dimensional non-rotating BTZ black hole. We obtain closed
expressions for it.}

\section{Introduction}


It is well-known that the energy-momentum tensor in Quantum Field
Theory in curved spacetime is a subtle issue. This is principally
due to the divergences which occur when the expectation value of
it in a certain state is calculated, see for example
\cite{ndBpcwD82} for an extensive discussion. However, since it
contains important physical information of the field it is worth
trying to calculate it. It turns out that in 1+1 dimensions most
of the difficulties can be removed and it is possible to obtain
closed expressions for it \cite{pDav77}. In this paper we will
exploit this fact and will calculate this quantity for the 1+1
dimensional non-rotating BTZ black hole\footnote{It is worth
mentioning that there exist the rotating BTZ black hole in 1+1
dimensions too, see for example \cite{aAchumOrt93} and
\cite{dGrurMc07}.}. Even though a great amount of work on
calculating the expectation value of the energy-momentum tensor in
several two dimensional spacetimes has been done, see for example
references in \cite{pDav77} and \cite{rBalaFab98}, as far as we
know the study of this tensor without walls in the 1+1 dimensional
non-rotating BTZ black hole has not been done before\footnote{A
similar study of the energy-momentum tensor in the BTZ black hole
in 1+1 dimensions with walls has been done in \cite{ecVage03}. The
results of this work can be considered complementary to the
present work.}. The main motivation of this work is to fill this
gap in the literature. Also it is worth mentioning that although
the calculations are very simple interesting results are obtained
and closed expressions as well.

\section{The energy-momentum tensor in 1+1 dimensions}

In 1+1 dimensions the energy-momentum tensor is almost determined
by its trace. In what follows we give the basic formule for
calculating this quantity.

Let us consider the following metric
\begin{equation}\label{E:1e6}
ds^{2}=C(-dt^{2}+dx^{2})=-Cdudv,
\end{equation}
where $u=t-x$ and $v=t+x$. Since every 1+1 dimensional metric is
conformal to a 1+1 dimensional metric in Minkowski spacetime, the
metric (\ref{E:1e6}) is very general. The function $C$ in general
depends on both variables in the metric. In these circumstances
the expectation value of the trace of the energy-momentum is
\cite{pDav77}
\begin{equation}\label{E:2e6}
<T^{\mu}_{\mu}>=-\frac{R}{24\pi}=\frac{1}{6\pi}\left(\frac{C_{uv}}{C^{2}}-\frac{C_{u}C_{v}}{C^{3}}\right),
\end{equation}
where $R$ is the Ricci scalar and $C_{u}=\frac{\partial}{\partial
u}C$, etc. The last expression holds for the real scalar field.
The expectation value of the components of the energy-momentum
tensor in null coordinates is
\begin{equation}\label{E:3e6}
\langle
T_{uu}\rangle=-\frac{1}{12\pi}C^{1/2}\partial_{u}^{2}C^{-1/2}+f(u)
\end{equation}
\begin{equation}\label{E:4e6}
\langle
T_{vv}\rangle=-\frac{1}{12\pi}C^{1/2}\partial_{v}^{2}C^{-1/2}+g(v)
\end{equation}
where $f$ and $g$ are arbitrary functions of $u$ and $v$
respectively. These functions contain information about the state
with respect to which the expectation value is taken. The mixed
components are given by
\begin{equation}\label{E:5e6}
\langle T_{uv}\rangle=-\frac{CR}{96\pi}.
\end{equation}
Now let us apply these formulae to the 1+1 dimensional
non-rotating BTZ black hole.

\subsection{The energy-momentum tensor in the 1+1 dimensional non-rotating BTZ black hole}

The metric for the 1+1 dimensional non-rotating BTZ black hole can
be written in the form (\ref{E:1e6}) with
\begin{equation}\label{E:6e6}
C=N^{2}=\left(-M+\frac{r^{2}}{l^{2}}\right)
\end{equation}
and $r^{*}=x$ where $r^{*}$ is the tortoise like coordinate
defined by $\frac{dr^{*}}{dr}=C^{-1}$. The function $C$ can be
written as function of $r^{*}$ or $u$ and $v$ as follows
\begin{equation}\label{E:6e7}
C=\frac{M}{\sinh^{2}\kappa
r^{*}}=\frac{M}{\sinh^{2}\kappa\frac{(v-u)}{2}},
\end{equation}
where $\kappa=\frac{r_{+}}{l^{2}}$ is the surface gravity with
$r_{+}=l\sqrt{M}$.

Using the previous expression for $C$ we obtain
\begin{equation}\label{E:7e6}
\langle T_{uu}\rangle=-\frac{\kappa^{2}}{12\pi}+f(u)
\end{equation}
and
\begin{equation}\label{E:8e6}
\langle T_{vv}\rangle=-\frac{\kappa^{2}}{12\pi}+g(v).
\end{equation}

Using that
\begin{equation}\label{E:9e6}
T_{tt}=T_{uu}+2T_{uv}+T_{vv},
\end{equation}
\begin{equation}\label{E:10e6}
T_{xx}=T_{uu}-2T_{uv}+T_{vv},
\end{equation}
and
\begin{equation}\label{E:11e6}
T_{tx}=-T_{uu}+T_{vv}
\end{equation}
we obtain
\begin{equation}\label{E:12e6}
\langle
T_{tt}\rangle=-\frac{\kappa^{2}}{6\pi}+\frac{CR}{48\pi}+f(u)+g(v),
\end{equation}
\begin{equation}\label{E:13e6}
\langle
T_{xx}\rangle=-\frac{\kappa^{2}}{6\pi}-\frac{CR}{48\pi}+f(u)+g(v)
\end{equation}
and
\begin{equation}\label{E:14e6}
\langle T_{tx}\rangle=g(v)-f(u).
\end{equation}

From the last three expressions it follows that
\begin{equation}\label{E:15e6}
\langle {T^{\mu}}_{\nu}\rangle=A+B+D
\end{equation}
where
\begin{equation}\label{E:16e6}
A=\frac{\kappa^{2}}{6\pi C}\left( {\begin{array}{cc}
 1 & 0  \\
 0 & -1  \\
 \end{array} } \right)
\end{equation}
\begin{equation}\label{E:17e6}
B=-\frac{R}{48\pi}\left( {\begin{array}{cc}
 1 & 0  \\
 0 & 1  \\
 \end{array} } \right)
\end{equation}
\begin{equation}\label{E:18e6}
D=\frac{1}{C}\left( {\begin{array}{cc}
 -f(u)-g(v) & f(u)-g(v)  \\
 g(v)-f(u) & f(u)+g(v)  \\
 \end{array} } \right).
\end{equation}

If we choose $f(u)=g(v)=0$ we obtain the analogue of the Boulware
state in Schwarzschild spacetime which is singular at the horizon
($C=0$). However we can also obtain the analogous of the
Hartle-Hawking state which is regular in both the future and the
past horizons. The value of $f(u)$ and $g(v)$ can be obtained if
we express the energy-momentum tensor in Kruskal like coordinates,
U and V. In these coordinates the energy momentum tensor is
\begin{equation}\label{E:19e6}
\langle T_{UU}\rangle=\frac{1}{U^{2}}\langle
T_{uu}\rangle=\frac{1}{U^{2}}\left(\frac{f(u)}{\kappa^{2}}-\frac{1}{12\pi}\right),
\end{equation}
\begin{equation}\label{E:20e6}
\langle T_{VV}\rangle=\frac{1}{V^{2}}\langle
T_{vv}\rangle=\frac{1}{V^{2}}\left(\frac{g(v)}{\kappa^{2}}-\frac{1}{12\pi}\right),
\end{equation}
and
\begin{equation}\label{E:21e6}
\langle T_{UV}\rangle=0.
\end{equation}
Hence we demand that $f(u)=g(v)=\frac{\kappa^{2}}{12\pi}$,
although it is not the only possibility. We could choose functions
which close to the horizon are have the values
$\frac{\kappa^{2}}{12\pi}$ and a different value far from it. So
$f$ and $g$ constants are not the only possibility. It is worth
pointing out that there is no natural analogue of the Unruh
vacuum, since this would lead us to have no conservation of
energy-momentum at infinity. Because of the timelike nature of
this boundary we must impose no net flux of momentum at infinity
and the Unruh state would violate this condition.

Now let us explain the physical interpretation for the functions
$f(u)$ and $g(v)$. From (\ref{E:18e6}) we see that these functions
are related with the flux of energy-momentum in the spacetime. For
example, for the Boulware state we have no flux of momentum at
infinity since $C$ goes to infinity there and kills this flux
momentum for any functions, however for this state we have flux of
energy-momentum at points in the exterior of the black hole. So in
a sense the state is not static. Whereas for the Hartle-Hawking
state there is no flux at no point of the black hole, so we have
a, let us say, static state, we just have energy density and
pressure at every point of the spacetime. So the functions $f(u)$
and $g(v)$ control the nature of the state on the black hole.

Also it is interesting to write this tensor in an orthonormal
frame. This can be done by introducing two-beins. The appropriate
orthonormal frame is given by
\begin{equation}\label{E:22e6}
{e^{a}}_{t}=(N, 0)
\end{equation}
and
\begin{equation}\label{E:23e6}
{e^{b}}_{r}=(0, N),
\end{equation}
where $a$ and $b$ are indexes associated with the orthonormal
frame. In these circumstances the energy-momentum tensor is given
by
\begin{equation}\label{E:24e6}
\langle T^{ab}\rangle=E+F+G
\end{equation}
where
\begin{equation}\label{E:25e6}
E=-\frac{\kappa^{2}}{6\pi C}\left( {\begin{array}{cc}
 1 & 0  \\
 0 & 1  \\
 \end{array} } \right)
\end{equation}
\begin{equation}\label{E:26e6}
F=\frac{R}{48\pi}\left( {\begin{array}{cc}
 1 & 0  \\
 0 & -1  \\
 \end{array} } \right)
\end{equation}
\begin{equation}\label{E:27e6}
G=\frac{1}{C}\left( {\begin{array}{cc}
 f(u)+g(v) & f(u)-g(v)  \\
 f(u)-g(v) & f(u)+g(v)  \\
 \end{array} } \right)
\end{equation}
From this expression we see that for the Hartle-Hawking state
\begin{equation}\label{E:28e6}
\langle T^{ab}\rangle=\frac{R}{48\pi}\left( {\begin{array}{cc}
 1 & 0  \\
 0 & -1  \\
 \end{array} } \right).
\end{equation}
Since in the present case\footnote{This is because locally the
non-rotating BTZ black hole is AdS spacetime.}
$R=-\frac{2}{l^{2}}$, then the energy density and the pressure are
given respectively by
\begin{equation}\label{E:28e6}
\rho=-\frac{1}{24l^{2}\pi}
\end{equation}
and
\begin{equation}\label{E:29e6}
p=\frac{1}{24l^{2}\pi}.
\end{equation}
Hence the radius $l$ determines the properties of the field in the
1+1 dimensional non-rotating BTZ black hole. It is clear from
(\ref{E:28e6}) that there is no flux of momentum at infinity
however there is a different from zero energy density at every
point of the spacetime.

It is also interesting to look at the semiclassical Einstein field
equations. It is well known that in 1+1 dimensions the Einstein
tensor vanishes identically, so there are no Einstein equations.
In particular the Einstein field equations with cosmological
constant in vacuum are impossible, however we now show that a kind
of Einstein field equations with cosmological constant make sense
when the right hand side of them is taken to be the expectation
value of the energy-momentum tensor we have found.

If we write the semiclassical Einstein field equations as
\begin{equation}\label{E:30e6}
R_{\mu\nu}-\frac{1}{2}Rg_{\mu\nu}+\Lambda g_{\mu\nu}=\langle
T_{\mu\nu}\rangle,
\end{equation}
with $\Lambda<0$, then in 1+1 dimensions the first two terms of
the left hand side vanish identically and we are left with
\begin{equation}\label{E:30e6}
\Lambda g_{\mu\nu}=\langle T_{\mu\nu}\rangle.
\end{equation}
But according to our expressions for the energy-momentum tensor
this equality can be satisfied if and only if
\begin{equation}\label{E:31e6}
g_{\mu\nu}=\frac{CR}{48\pi\Lambda}\left( {\begin{array}{cc}
 1 & 0  \\
 0 & -1  \\
 \end{array} } \right)
\end{equation}
which is no other thing than the metric for the BTZ black hole in
1+1 dimensions, scaled by an overall factor and multiplied by
minus one. This could be interpreted as the metric inside the
horizon. Hence we could say that the back reaction shifted the
horizon by making it bigger. It is interesting to note that if the
sign would be opposite then there would not be change in the
geometry. Since if we take this metric as the starting point for
calculating the expectation value of the energy-momentum tensor we
would find the same values as previously \cite{rBalaFab98}. In
this second scenario the BTZ metric would be stable under back
reaction effects. The expression (\ref{E:31e6}) and the previous
analysis could be suggestive and take
$\Lambda=\frac{R}{48\pi}=-\frac{1}{24\pi l^{2}}$ in two
dimensions, so we would have an analogous expression for two
dimensions to that of other dimensions where
$\Lambda=\frac{d(1-d)}{2l^{2}}$ and $d+1$ is the dimension of the
spacetime.

The previous discussion should be taken with care and just as an
indication of the possible scenarios, since there is no Einstein
equations in 1+1 dimensions. A more natural thing to do would be
to plug in the expectation value of the energy-momentum we have
found into a theory of gravity in 1+1 dimensions and see how it is
modified.\vspace{0.5cm}


Acknowledgments: I thank my supervisor, Dr. Bernard S. Kay, his
guidance and several conversations during this work. Also I thank
Dr. Daniel Grumiller for helpful comments about this work. The
corrections to a previous version of this work were carried out at
Centro Universitario UAEM Valle de Teotihuacan, Axapusco, State of
Mexico, Mexico.

This work was carried out with the sponsorship of CONACYT Mexico
grant 302006.

\end{document}